\documentclass{ws-ijmpd}
\usepackage[super,compress]{cite}
\usepackage[dvips,breaklinks]{hyperref}
\usepackage{graphicx}
\usepackage{amssymb}
\usepackage{amsmath}
\usepackage{bbding}
\usepackage{color}

\newcommand{\beq}{\begin{equation}}
\newcommand{\eeq}{\end{equation}}
\newcommand{\ignore}[1]{}
\newcommand{\be}{\begin{equation}} \newcommand{\ee}{\end{equation}}
\newcommand{\bea}{\begin{eqnarray}} \newcommand{\eea}{\end{eqnarray}}
 \renewcommand{\bf}{\textbf}

\begin{document}
\title{Polarization Alignment  in JVAS/CLASS flat spectrum radio surveys }
\author{Prabhakar Tiwari and Pankaj Jain }
\address{Department of Physics, Indian Institute of Technology, Kanpur - 208016, India}
%\date{29 May,2012}
%\pagerange{\pageref{firstpage}--\pageref{lastpage}} %\pubyear{2012}

%\label{firstpage}
\maketitle

\begin{abstract}
We present a detailed statistical analysis of the 
alignment of polarizations of 
radio sources at high redshift.
This study is motivated by  
the puzzling signal of alignment of polarizations from distant 
quasars at optical 
frequencies. 
 We use a coordinate invariant measure of dispersion to test for
alignment of polarizations of widely separated sources. The data from 
JVAS/CLASS 8.4-GHz surveys 
are used for our study.
We find that the data set with polarization flux greater than 1 mJy 
shows a significant
signal of alignment at 
distance scales of order 150 Mpc. The significance of alignment decreases 
as we go to larger distances. 
In contrast the data set with polarization
flux less than 1 mJy does not show significant alignment at short 
distances. However this set, as well as the full data sample, shows a
very significant signal at large distances of order 500-850 Mpc. 
Due to the presence
of relatively large error in the 
low polarization sample, this signal needs to be tested
further with more refined data. 
 Surprisingly, we also find that, at large distances,  
high polarization sample shows anomalously large 
scatter in comparison to random data sets, generated by shuffling 
data polarizations.  
We suggest that this might arise since the distribution of polarizations
is not exactly uniform.
We also study the signal by imposing a cut on the error in
polarization. We find that the significance of alignment increases with
decrease in fractional error. 
We are unable to attribute our results to known sources of bias.
We discuss a possible physical explanation of our results.

\end{abstract}
%%%%%%%%%%%%%%%%%%%%%%%%%%%%%%%%%%%%%%%%%%%%%%%
\section{Introduction}
%%%%%%%%%%%%%%%%%%%%%%%%%%%%%%%%%%%%%%%%%%%%%%%

The Big Bang model assumes that the Universe is homogeneous and isotropic 
on large distance scales. However there currently exist several observations
which appear to violate this basic assumption. In particular the radio
polarizations from radio galaxies 
appear to show a large scale dipole pattern across the
sky \cite{Birch:1982,Kendall:1984,Jain:1998r}. 
The possibility that the signal observed by \cite{Birch:1982} might arise due to
bias was raised by \cite{Phinney:1983}. 
The signal 
was dismissed by \cite{Bietenholz:1984} who found that it is not present
in a larger data set. However \cite{Jain:1998r} argued that the signal is present
if we consider all the radio sources for which the relevant information, i.e.
the polarization position angle and the galaxy orientation angle, is available in the literature. The relevant observable in this study was the difference
of these two angles, which is independent of the coordinate system.  
It is also interesting that the dipole axis found in \cite{Jain:1998r} aligns
closely with the Cosmic Microwave Background Radiation (CMBR) dipole axis.
Furthermore a recent study 
finds a dipole anisotropy in the brightness of radio sources which is 
much larger than what is predicted by the Doppler effect due to local motion \cite{Singal:2011}. This also 
indicates a dipole axis which may be of cosmological origin and well aligned with the CMBR
dipole.

The optical polarizations from quasars also show alignment over 
cosmologically large distances 
\cite{Hutsemekers:1998,Hutsemekers:2000fv,Jain:2003sg}. The distance scale of these correlations is of the
order of Gpc \cite{Jain:2003sg}. Furthermore the Cosmic Microwave 
Background Radiation (CMBR) data shows several features 
\cite{Tegmark:2004,Eriksen:2004,Ralston:2004,Land:2005,Kim:2010,Samal:2008,Samal:2009} 
which are not
consistent with Big Bang cosmology. We point out that the WMAP science team has argued that 
some of the claimed anomalies in CMBR data set may arise due to a posteriori choice of statistics
to test for a particular effect \cite{Bennett:2011}. In other words one notices a particular
odd feature in the data and then devises a statistic to test its significance. Such a procedure
is likely to overestimate the significance of the detected anomaly. In view of this it is
extremely interesting that the alignment axis of CMBR quadrupole and octopole \cite{Tegmark:2004},
radio dipole axis \cite{Jain:1998r} and the two point correlations in the optical polarizations
\cite{Hutsemekers:1998,Hutsemekers:2000fv,Jain:2003sg} all align very 
closely with the 
CMBR dipole axis \cite{Ralston:2004} and point roughly in the direction of the Virgo cluster. The maximum angular separation between any of these two  
axes is found to be $20^o$ \cite{Ralston:2004}. 
If we remove the CMB dipole, the remaining
axes align even more closely with one another. 
Furthermore there have been claims
of violation of isotropy in cluster peculiar velocities \cite{Kashlinsky:2009t,Kashlinsky:2009r}
 and galaxy surveys \cite{Itoh:2010}.
Remarkably the cluster peculiar velocities also indicate a direction close 
to the CMB dipole if we include the highest redshift data. 
Finally we mention that spiral galaxies reveal an interesting signal of parity 
violation \cite{Longo:2011}. 

There exist many attempts to theoretically 
explain these observations, most of which assume violation of the cosmological
principle. An interesting possibility, which is completely consistent
with Inflationary Big Bang model, is that the Universe was 
inhomogeneous and anisotropic at very early stage, before the epoch 
of inflation. It evolves into a homogeneous and isotropic
de Sitter space-time during inflation. 
It has been shown that, for a wide class of
models, there exists a parameter range such that the modes generated during
this early phase can re-enter the horizon much before the current era
and hence can affect present observations \cite{Aluri:2012}. These can, in principle, generate
the observed anisotropies. 

In \cite{Jackson:2007}, the authors 
compiled a catalogue of radio polarizations from distant radio galaxies.
 Motivated by
the observed linear polarization alignment in quasar data 
\cite{Hutsemekers:1998,Hutsemekers:2000fv} at visible wavelengths, 
\cite{Joshi:2007} studied the possibility of a similar effect at radio frequencies
using this catalogue. No significant effect was detected. 
The alignment in optical polarizations 
is seen over cosmologically large distances of order Gpc 
\cite{Hutsemekers:1998,Hutsemekers:2000fv,Jain:2003sg} 
and the phenomena is very puzzling. There are several  possible models
\cite{Hutsemekers:1998,Hutsemekers:2000fv,Jain:2002vx,Hutsemekers:2005iz,Payez:2008pm,Urban:2011,Ciarcelluti:2012}, 
which aim to provide an interpretation of this alignment.
One possible explanation is the large scale correlations
in the intergalactic magnetic field
\cite{Agarwal:2008ac}. The intergalactic magnetic field may be seeded 
in the early Universe
\cite{Subramanian:2003sh,Seshadri:2005aa,Seshadri:2009sy}
and presence of large scale correlations in this field cannot be ruled  
out within the Big Bang cosmological model \cite{Agarwal:2008ac}.
We still require a mechanism for how such a magnetic field generates
large scale correlations in the optical polarizations. One possibility 
is that this is caused by mixing of photons with hypothetical pseudoscalars
\cite{Jain:2002vx,Hutsemekers:2005iz,Payez:2008pm,Agarwal:2008ac}. This effect is frequency dependent and is much smaller at
radio frequencies \cite{Jain:2002vx}. Hence it may be consistent with absence or reduced alignment
effect in radio polarizations. 
 There have also been other proposals to explain this effect
\cite{Urban:2011,Ciarcelluti:2012}. 

In the present paper we analyze the radio data in order to study possible
alignment of radio polarizations. 
 Here we extend the work of \cite{Joshi:2007}
 by considering the fact that
polarization angles depend on the coordinate system used. In
order to properly analyze the presence or absence of large scale
alignment one needs to define a coordinate invariant statistics
\cite{Jain:2003sg}. The basic point is that, in order
to compare two polarization angles on the surface of the celestial sphere,
one needs to parallel transport one of them to the position of the
second along the great circle joining the two points. The contribution
due to parallel transport may be negligible in most cases, if the
two points are separated by a small distance. However it becomes very
important if we are testing alignment over large distances.
In \cite{Joshi:2007}, the authors restricted
their analysis to data which has polarization flux density greater than
1 mJy in order to select data which only contains sources with significant
polarization detection. Here we study the signal both
in the high polarization and the low polarization data sample.
Furthermore we study data 
after imposing a cut on the fractional error in polarization
since sources with small error are likely to be most reliable.

A potential source of bias is the error in the removal of residual
instrumental polarization \cite{Jackson:2007,Joshi:2007}. This can lead
to large scale correlations in polarizations even when none are present. 
It is clear that this effect will dominate for sources which have 
low polarization flux. Hence one can evaluate its contribution 
by focussing on data
with low polarizations. Another source of error  
is the positive bias \cite{Simmons:1984,Jackson:2007} that arises 
in the degree of polarization. This arises since the degree of polarization
depends on the squares of Q and U and  always acquires a positive value. 
However this cannot 
 affect the alignment
statistics which are based only on the linear polarization angle.
We also point out that errors in polarization position angle (PA) calibration
can reduce the true significance of alignment that might be present in data. 
This will lead to a systematic error in the observed PA which may be 
different in different runs \cite{Joshi:2007}.  
Hence, this can mask 
an alignment effect present in the data, but should not 
introduce spurious alignments.

%%%%%%%%%%%%%%%%%%%%%%%%%%%%%%%%%%%%%%%%%%%%%%%
\section {Data Selection}
%%%%%%%%%%%%%%%%%%%%%%%%%%%%%%%%%%%%%%%%%%%%%%%

We use data available in the catalogue produced 
by \cite{Jackson:2007}. It contains a
total of 12743 core dominated flat spectrum radio sources
and lists their
angular positions and the Stokes $I$, $Q$ and $U$ parameters.
Since the redshift of most of these sources is unknown,
we assume that these
sources are roughly at the same redshift, equal to unity. 
The input observable for
the alignment study is the polarization angle.
The calibration methods and the catalogue production has been discussed in detail in \cite{Jackson:2007}.

In \cite{Joshi:2007}, the authors imposed a cut on 
polarization flux density to include only sources with polarization
 flux greater than 1 mJy. 
We also impose this cut on the data.   
The resulting set contains a total of 4400 sources.
We shall refer to this as set $1$. We also consider the    
data sample with polarization flux greater than 
2 mJy (set 2), less than 1 mJy (set 3) and less than 0.5 mJy (set 4).
These contain 2468, 8342 and 5283 sources 
respectively.
We note that the sources lie 
dominantly in the Northern hemisphere. Furthermore there are very few 
sources along the Galactic plane. 

\begin{figure}[!t]
    \includegraphics[width=6.0in,angle=0]{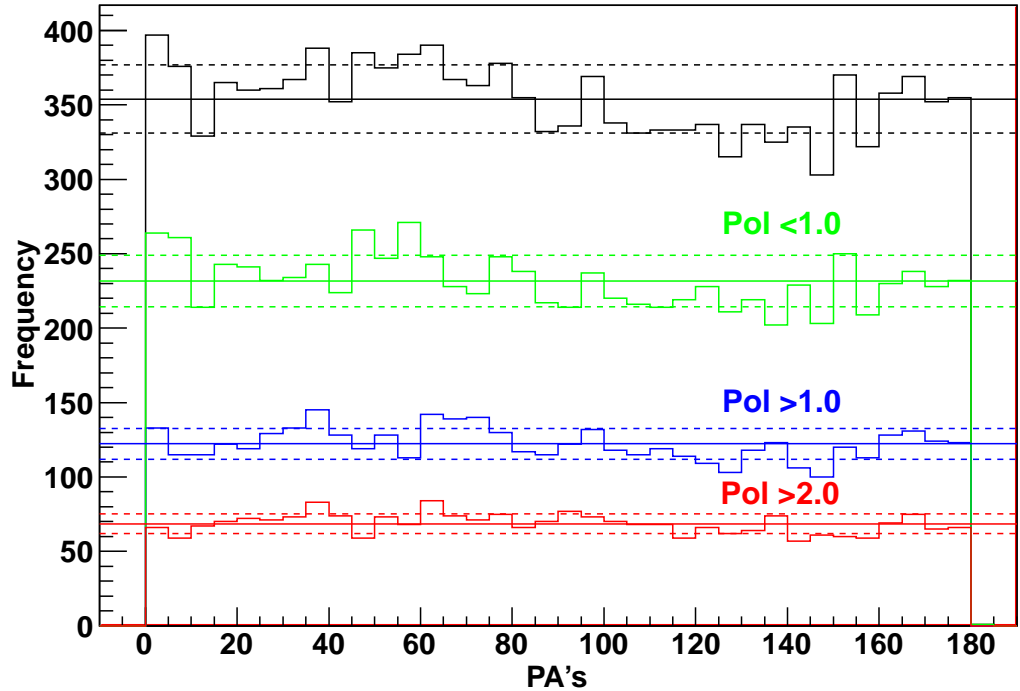}
    \caption{
  The distribution of polarization position angles
(PAs) for the complete data set (upper curve), data with polarization flux
less than 1 mJy, greater
than 1 mJy and greater than 2 mJy, as indicated in the plot. 
The mean and $\pm 1 \sigma$ lines for each
distribution are also shown.
}
\label{fig:PAs}
\end{figure}

The distribution of polarization angles for the data sets with different
cuts is shown in Fig. \ref{fig:PAs}. 
We notice that the distribution in all cases 
shows a small bump for 
$PA<90^o$ and a corresponding dip for $PA>90^o$. This trend is more
prominent for data with small polarizations. 
 We may quantify this non-uniformity by making a fit with the von-Mises
distribution,
\begin{equation}
f(\theta,\kappa) = {1\over 2\pi I_0(\kappa)} \exp{[\kappa(\theta-\theta_0)]}
\end{equation}
where $I_0$ is the modified Bessel function of order 0. In making our 
fit we set $\theta=2 PA$ \cite{Sarala:2001}. The null hypothesis is that
the distribution is uniform, i.e.
$\kappa = 0$. The best fit values of $(\kappa, \theta_0/2)$ are found 
to be $(0.064,40^o)$, $(0.070,46^o)$ and $(0.063,37^o)$ for the
full data, $Pol>1$ and $Pol<1$ respectively. Using likelihood analysis, we
find that the fit is significant at 4.7 $\sigma$, 3.6 $\sigma$ and
2.8 $\sigma$ respectively for these cases.  
Hence the distribution shows a significant deviation from uniformity.
We do not understand the reason for this non-uniformity.
In any case, as we discuss in the next
section, the non-uniformity in distribution does not affect our
results. 

%%%%%%%%%%%%%%%%%%%%%%%%%%%%%%%%%%%%%%%%%%%%%%%
\section{Statistical Procedure}
%%%%%%%%%%%%%%%%%%%%%%%%%%%%%%%%%%%%%%%%%%%%%%%

All astronomical observations are made on the hypothetical  celestial sphere and any directional
measurement on this sphere corresponds to a particular coordinate system.
For example, the polarization angles are measured in a local frame, formed
by two unit vectors
 $\hat{\phi}$ and $\hat\theta$. These unit vectors
depends on the coordinate system used, i.e. they depend on which direction we
choose as our North Pole, and hence one cannot directly
compare vectors at different positions on the celestial sphere. The proper
procedure to compare such 
vectors is to transport one of them
to the position of the second along the geodesic joining the two positions.
A detailed procedure has been discussed in
\cite{Jain:2003sg} and we follow this procedure for 
comparing polarization
angles of two different sources. 

We next define statistics in order to quantify the alignment 
\cite{Hutsemekers:1998,Jain:2003sg}.
This requires a measure of the difference of polarization angles
at different sites. Let the polarization angles at site $i$ and $j$
be $\psi_i$ and $\psi_j$ respectively. 
We associate the unit vectors $\hat v_i=[\cos(2\psi_i),\sin(2\psi_i)]$ 
and $\hat v_j=[\cos(2\psi_j),\sin(2\psi_j)]$ 
with these polarization angles.
The factor of two is required since a polarization
angle $\psi=0$ is identified with $\psi=180^o$. The variable $2\psi$
takes values over the entire range $0$ to $360^o$. This is discussed
in more detail by \cite{Ralston:1999}.  
A convenient measure of alignment between vectors $\hat v_i$ and
$\hat v_j$ is given by their dot product,  
$\hat v_i\cdot \hat v_j=\cos(2\psi_i-2\psi_j)$.  
This cosine measure takes into account the angular 
nature of polarizations and can also be motivated 
by the von Mises distribution, relevant for data on a circle
\cite{Fisher:1993}. 
Taking into account the contribution due to parallel transport,
the dot product becomes $\cos[2(\psi_i+\Delta_{i\rightarrow j})-2\psi_j]$,
 where $\Delta_{i\rightarrow j}$ is the contribution due to 
parallel transport from site $i$
to site $j$ along a great circle joining these two points \cite{Jain:2003sg}.
An alternate measure of alignment is $|(\pi/2)-|\psi_i-\psi_j||$ 
\cite{Hutsemekers:1998}, where the angles are in radians. 

Using the measures discussed above, we define three statistics, 
which can be used to test for alignment. We first define the statistic,
$S_D$, which has been used in earlier studies 
\cite{Hutsemekers:1998,Jain:2003sg}. 
Consider the source $k$ along with it's $n_v$ nearest neighbours.
A useful measure of alignment of polarizations 
in the neighbourhood of site $k$ is given by, 
\bea
d_{k} = \frac{1}{n_{v}} \sum_{i=1}^{n_{v}} \cos[2(\psi_{i}+\Delta_{i\rightarrow
k}) - 2{\bar\psi}_{k})].
\label{eq:dispersion}
\eea
Here $\psi_i$ are the polarization angles of the nearest neighbours of the $k^{th}$ site
and the factor $\Delta_{i\rightarrow k}$ arises due to 
parallel transport from $i\rightarrow k$. The sum in Eq. \ref{eq:dispersion}
includes the site $k$ also.
Essentially the polarizations of the nearest neighbours are first
parallel transported
to the site $k$ and then tested for alignment.  
We next maximize $d_k$ as a function of ${\bar\psi}_k$. 
The resulting value of ${\bar\psi}_k$ is interpreted as the mean polarization angle at the site $k$ and the 
corresponding maximum of $d_k$ takes on higher values for data 
with lower dispersions and vice versa.
The statistic may now be defined as \cite{Hutsemekers:1998,Jain:2003sg},
\bea
S_{D} = \frac{1}{n_{s}} \sum_{k=1}^{n_{s}} d_{k}\big|_{\rm max},
\label{eq:statistics}
\eea
where $n_s$ is the total number of samples in the data.
For example, for set 1, $n_s=4400$.
A large value of $S_{D}$ indicates a strong alignment between polarization vectors. 

An alternate statistic that we also use is defined as,
\bea
S'_{D} = \frac{1}{n_{s}} \sum_{k=1}^{n_{s}} d'_k 
\label{eq:statisticsp}
\eea
where
\bea
d'_{k} = \frac{1}{n_{v}} \sum_{i=1}^{n_{v}} \cos[2(\psi_{i}+\Delta_{i\rightarrow
k}) - 2{\psi}_{k})].
\label{eq:dispersionp}
\eea
Note that here $\psi_k$ is the
polarization angle at the site $k$. Hence, if we ignore the correction
due to parallel transport, $d'_k$ is simply the sum of cosines of the 
differences of twice
the polarization angle at site $k$ with it's nearest neighbours.
A third statistic we use is defined as,
\bea
Z_{D} = \frac{1}{n_{s}} \sum_{k=1}^{n_{s}} d''_k 
\label{eq:statisticspp}
\eea
with
\bea
d''_{k} = \frac{1}{n_{v}} \sum_{i=1}^{n_{v}} 
\Big|(\pi/2)-\big|(\psi_{i}+\Delta_{i\rightarrow k}) - {\psi}_{k}\big|\Big|.
\label{eq:dispersionpp}
\eea
In our analysis we mostly use statistic $S_D$ since this was also used earlier
in the analysis of optical data \cite{Hutsemekers:1998,Jain:2003sg}.
We use the statistics $S_D'$ and $Z_D$ to confirm the results obtained
by $S_D$.
 
The significance of alignment in the data set is computed as follows.
We first compute the  statistic
for a given number of nearest neighbours, $n_v$, of any source. 
This statistic is compared with the result for a large number of random
samples, which are generated by two different procedures. In {\it Procedure I}, 
the random samples are generated by
shuffling all the PAs among different sources. In {\it Procedure II}, these 
are generated from a uniform distribution of polarization angles.
Procedure I tests for local alignment of polarizations given the 
polarization distribution of data. Procedure II, in contrast, tests for
alignment relative to a uniform distribution.  
In most of our simulations we use
a total of 1000 random samples for a given number of nearest neighbours. 
The probability or P-value 
that the alignment seen in data might arise as a random
fluctuation is equal to the number of random samples which show
a larger  
value of statistic in comparison to the real data divided by the 
total number of random samples.  
Almost all the P-values quoted in results are 
computed directly by this procedure. 
In some cases, a reliable estimate of P-value was obtained only
after enhancing the number of random samples since we did not find any 
random sample exceeding the statistic of the real data. 
In some cases, the P-value was found to be too small and a direct estimate
was impractical. In this case it was estimated by computing the sigma value.
Here we assume that the distribution
of random samples is approximately Gaussian and  
determine it's mean
and standard deviation. This provides an estimate of the
 sigma values, which can be used to compute the P-values. 
We find that this
 slightly underestimates the significance in comparison
to a direct computation. However the difference
is found to be negligible for our purpose. 

The distribution of $PAs$ is slightly non-uniform, as shown in 
Fig. \ref{fig:PAs}. This non-uniformity does not affect our results as long
as we use Procedure I for generating the random samples. This
is because, in this case, the random samples are generated from
the same distribution as the real data. However we shall get somewhat
larger significance if we use Procedure II, which generates random
samples from the uniform distribution. Procedure I really tests
the alignment of sources, within a local neighbourhood, independent 
of their distribution. Hence, for example, it would yield a null
result for alignment if all the sources have the same PAs.  
This is discussed further in section 5. 

\begin{figure}[!t]
\includegraphics[width=3.5in,angle=0]{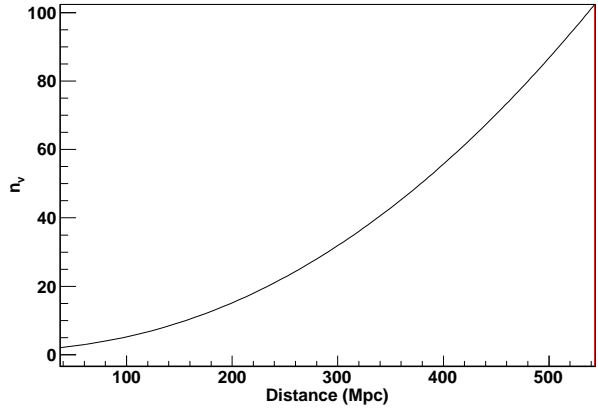}
  \caption{
 The mean distance (Mpc) among sources as a function
of the number of nearest neighbours ($n_v$) for data 
with polarization flux greater than 1, i.e. set 1. 
}
\label{fig:nv_dis}
\end{figure}

The number of nearest neighbours, $n_v$, of any source are computed by assuming
that all the sources are located at the same redshift of 1.0. The redshift information
of these sources is not currently available and hence we make this assumption.
Essentially here we are ignoring the third dimension and determining 
the nearest neighbours only on the basis of the angular separations.
It is likely that in many cases this will lead to a wrong assignment 
of the set of nearest neighbours of a source. However this cannot
generate alignment in a data set if none is present. If a data set shows 
alignment then it will
affect the detailed numerical results. For example, let us assume that
the radio polarizations are aligned over a small distance scale of a few
Mpc, but show no alignment over larger distances. Our nearest neighbour
assignment may include some sources which are infact much further away. 
It is clear that these sources which are mis-identified as nearest neighbours
will only add noise to the signal and reduce the significance of alignment.  
In Fig. \ref{fig:nv_dis},
we show the relationship between the number  
of nearest neighbours and the mean comoving distance from a source within which
these nearest neighbours reside. Here the mean is taken over the entire 
sample.  
 We have assumed the standard Lambda Cold Dark Matter model 
for computing the comoving distance \cite{Weinberg:2008}. 

%\begin{figure}[!t]
%  \includegraphics[width=3.5in, angle=0]{radio_sources_full_b.eps}
% % \caption
%  \bf{Figure 1.}  The sky distribution of radio sources with 
%polarization flux greater than 0.5 mJy. The lines represent 
%the linear polarizations.
%  \label{fig:sources}
%  \end{figure}

\begin{figure}[!t]
    \includegraphics[width=3.5in,angle=0]{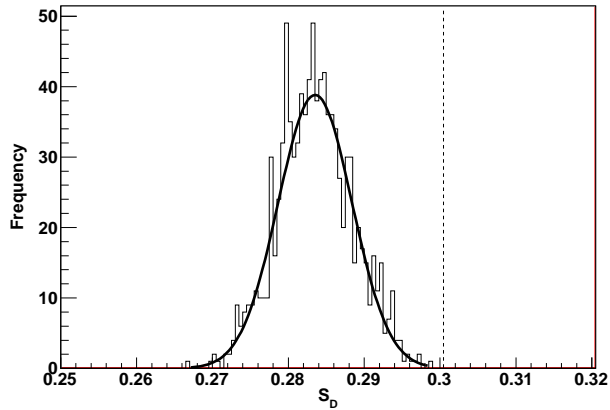}
    \caption{
 The distribution of statistic $S_D$ for data set 1 
  with the number of nearest 
neighbours, $n_v= 10$. The random samples are generated by shuffling the 
polarization angles among different sources (Procedure I). The Gaussian fit to the 
distribution is also shown. The dashed vertical line
shows the statistic for the real data set with $n_v=10$. 
}
\label{fig:stat_nv}
\end{figure}

\begin{figure}[!t]
    \includegraphics[width=4.5in,angle=0]{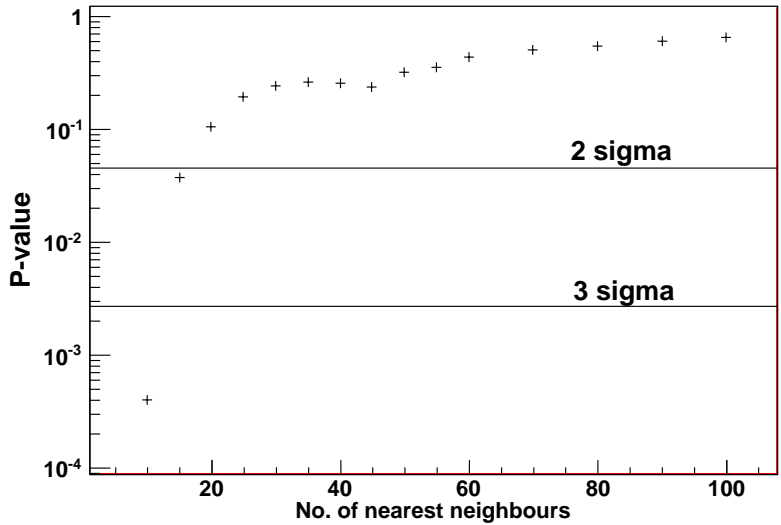}
    \caption{
The P-values, 
i.e. the significance of alignment, 
 as a function of the number of 
nearest neighbours, $n_v$, for
set 1. The 2 sigma and 3 sigma lines for reference. 
}
\label{fig:SD}
\end{figure}

\begin{figure}[!t]
%    \includegraphics[width=5.5in,angle=0]{statistics1.0_Oct19_30.eps}
%\newline
    \includegraphics[width=5.5in,angle=0]{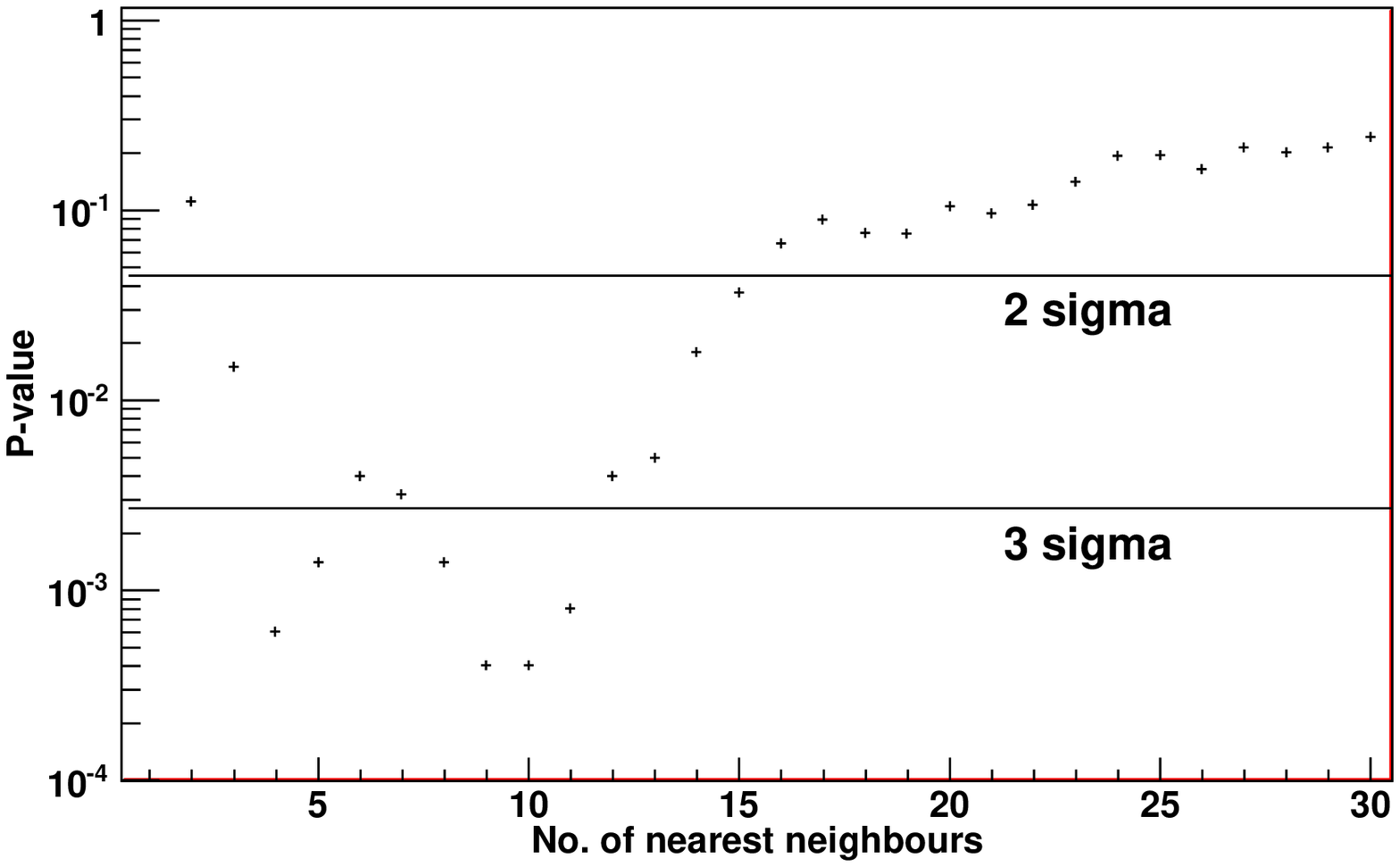}
    \caption{
The P-values 
(plusses) 
as a function of the number of 
nearest neighbours, $n_v$, for data set 1 using Procedure I. 
}
\label{fig:SD1}
\end{figure}

 %%%%%%%%%%%%%%%%%%%%%%%%%%%%%%%%%%%%%%%%%%%%%%%%%%%%%%%%%%%%
\section{Results}
\label{sc:results}
The distribution of statistic, $S_D$, for random samples is shown 
in Fig. \ref{fig:stat_nv} for
$n_v=10$. Here we use 1000 random samples generated by 
shuffling data for radio sources in set 1.
 The value of statistic for the real data  
set is also shown.  
We find that 
a Gaussian provides a good fit to the random sample distribution. 
The fit values for 1000 random samples 
 are found to be,
${\rm mean} = 0.2836$ and standard deviation=0.00495. If the number of
samples are reduced to 500 we again find a good fit with mean = 0.2834
and standard deviation=0.00492. 
We have explicitly verified that a direct evaluation of
the significance agrees well with that determined  
by the Gaussian fit. Hence the significance
we quote is reliable, despite the small number of nearest 
neighbours.

In Fig. \ref{fig:SD} 
significance of alignment, i.e. P-values, for data set 1
 a function of the number of nearest neighbours.  
In Fig. \ref{fig:SD1} we show the significance more clearly for  
small number of nearest neighbours.   
We do not find a significant 
alignment for $n_v>11$. This is in agreement
with the results obtained in 
\cite{Joshi:2007}. 
However for lower values of $n_v$ we find significant alignment,
with significance greater than 
3 sigmas, for several values of $n_v$. 
In \cite{Joshi:2007} this region was
never explored. The strongest signal is observed in the neighbourhood
of $n_v=10$. The mean comoving 
distance among sources for 10 nearest
neighbours is about 150 Mpc. 
Hence our results show that for polarization flux greater than 1 mJy, the
radio sources show alignment over distance scale of order 150 Mpc. 
This set is expected to be least contaminated with bias and hence
the alignment effect we observe is most likely of physical origin.
We point out that we 
do not find a significant signal of alignment for very small values
of $n_v$, as shown in Fig. \ref{fig:SD1}. A similar phenomenon was seen
even in the optical alignment. In this case also alignment was not seen
for very small $n_v$ \cite{Hutsemekers:1998}. This 
can be explained by the presence of large fluctuations at small 
$n_v$.  We discuss this further in section 6.

\begin{figure}[!h]
    \includegraphics[width=5.0in,angle=0]{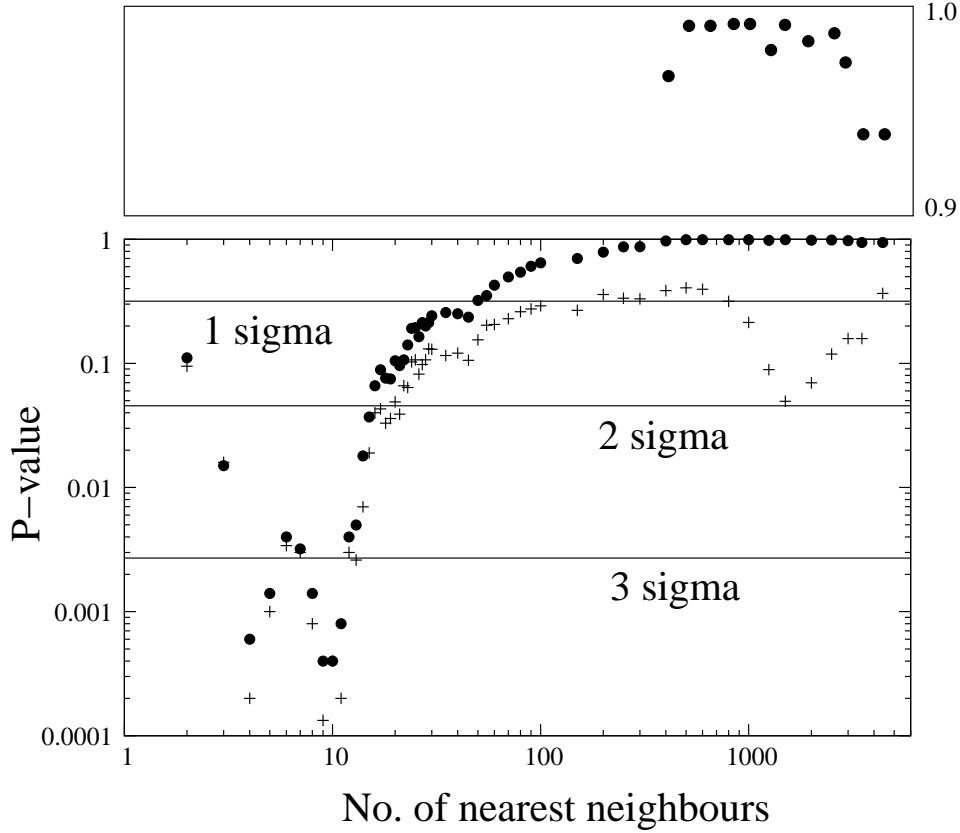}
    \caption{
The P-value  
as a function of the number of 
nearest neighbours, $n_v$, for data set 1 with statistic $S_D$
using Procedure I (filled circles) and Procedure II (plusses) 
for generating the random samples. 
The upper panel shows the expanded y-axis in 
the range $0.9$ to $1$
to show the points for $n_v\ge 500$ using Procedure I with better clarity. 
}
\label{fig:Pvalue_noshuffle}
\end{figure}

In Fig. \ref{fig:SD} we notice that, for large $n_v$, the significance of alignment continues
to decrease with increase in 
$n_v$. If, beyond a certain value of $n_v$, the polarizations were 
randomly aligned, 
we would have expected that the sigma value would fluctuate
about 0. In order to study this in more detail we have computed the
significance for larger values of $n_v$. 
The results for the P-values are shown in Fig. \ref{fig:Pvalue_noshuffle}
as filled squares. 
We find that the P-value continues to increase and rises 
above 0.95 beyond $n_v=500$. Equivalently, the sigma value
 falls below $-2$ beyond $n_v=500$. The P-value 
continues to increase till about $n_v=1000$ and then
starts decreasing.
 This implies that for $n_v$ between 500 and 1000,
the data set shows larger scatter in comparison
to a random sample. We have verified that this happens only for the 
actual data used. If instead we use a randomly generated sample, then
the sigma values indeed fluctuate about zero, as expected. This confirms
the correctness of our procedure and simulations. However surprisingly
the data sample shows unusual scatter for large $n_v$. 

In order to understand the behaviour at large $n_v$ 
better we generated the random 
samples directly without shuffling the real data. The distribution
of random samples is assumed to be uniform, i.e. constant as a function
of the polarization position angle.
The resulting P-values are also shown in Fig. \ref{fig:Pvalue_noshuffle}
as plusses. 
We find that, in general, this analysis yields higher significance 
in comparison to what
is obtained by shuffling the polarizations among different sources.
 At low $n_v$ we obtain 
a trend similar to that seen in Figs. \ref{fig:SD} and \ref{fig:SD1}. 
However at large $n_v$ we see a very different behaviour. The P-value
rises to about 0.4 at $n_v=500$. Beyond this it first falls and then again 
starts increasing. 
This analysis shows that the data set does not show unusual scatter
in comparison to a truly random sample. It is only with respect to
the shuffled sample that we obtain the anomalous result. This 
trend might be attributed to the non-uniformity of   
the distribution of polarizations in the real sample, as discussed in
section 2.

\begin{figure}[!h]
    \includegraphics[width=5.0in,angle=0]{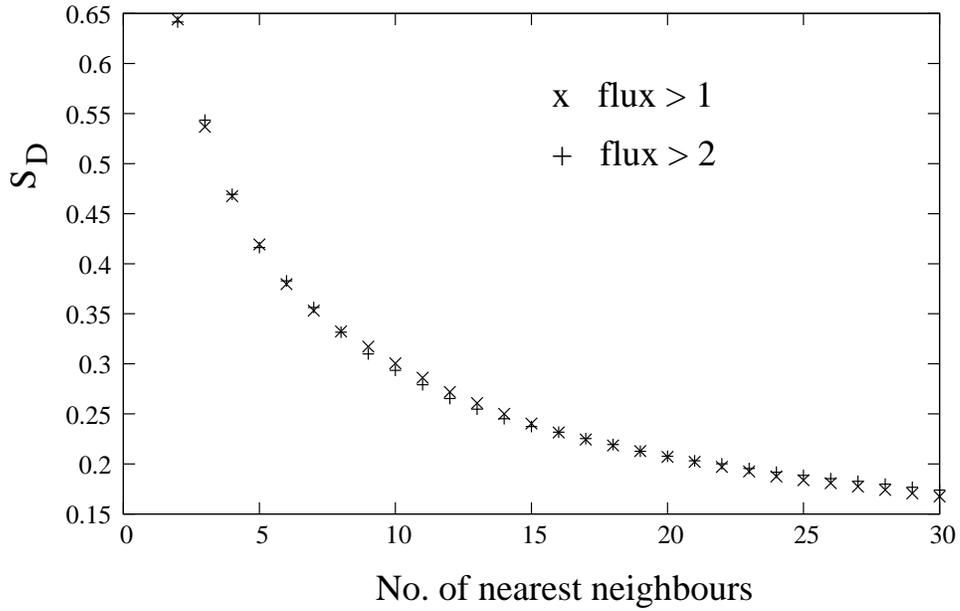}
    \caption{
The statistic $S_D$  
as a function of the number of 
nearest neighbours, $n_v$, for data set with polarization flux greater than
2 mJy (set 2). 
The results for set 1 are shown for comparison. 
}
\label{fig:SD_Polgt2}
\end{figure}

\begin{figure}[!h]
    \includegraphics[width=5.0in,angle=0]{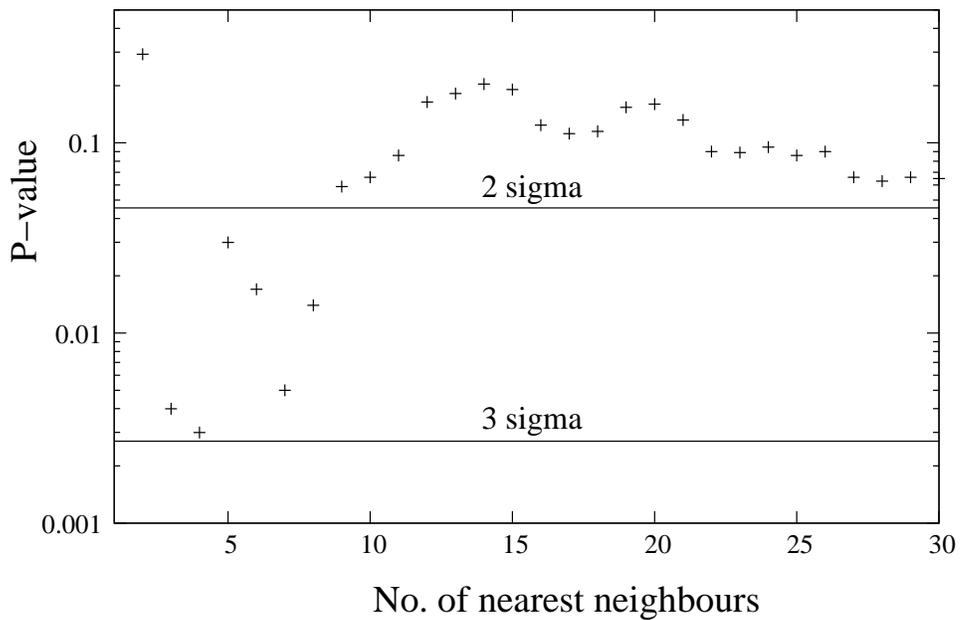}
    \caption{
The P-value  
as a function of the number of 
nearest neighbours, $n_v$, for data set 2. 
}
\label{fig:Pvalue_Polgt2}
\end{figure}

\begin{figure}[!t]
    \includegraphics[width=4.5in,angle=0]{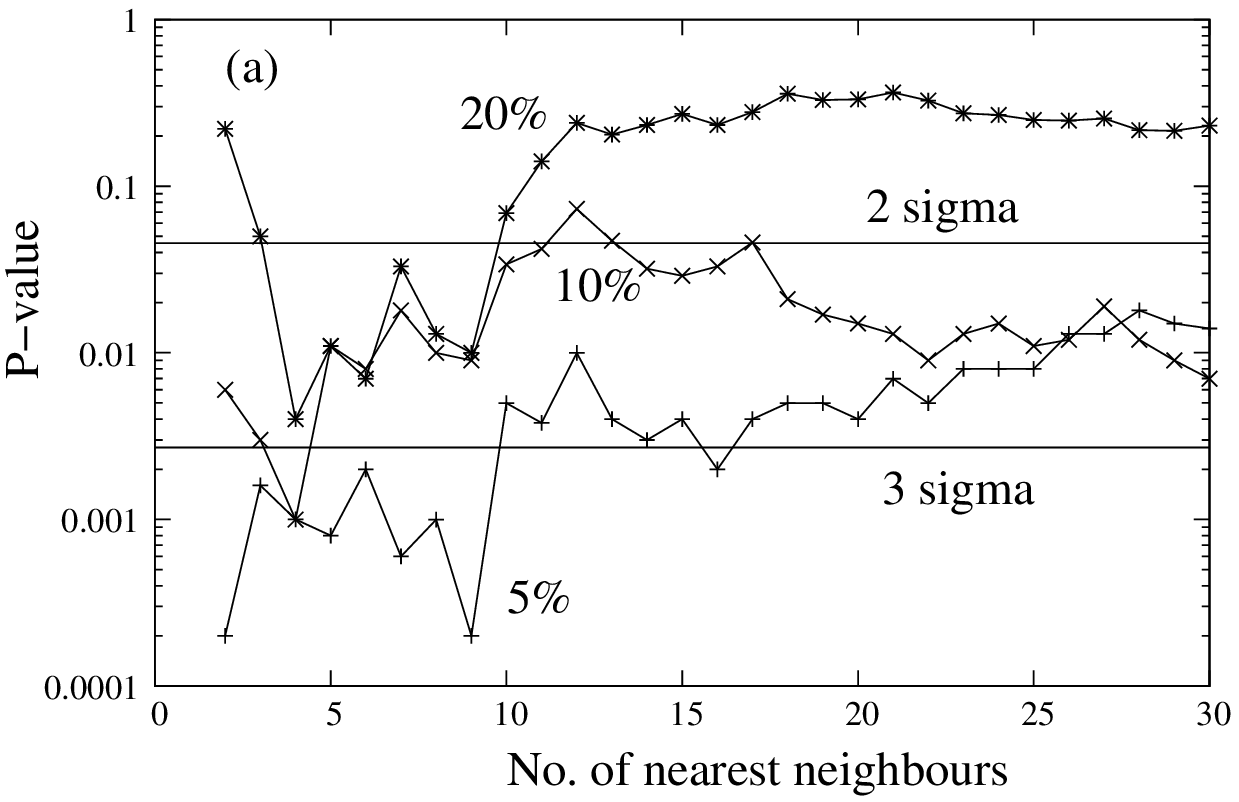}
\vskip 0.5 cm
    \includegraphics[width=4.5in,angle=0]{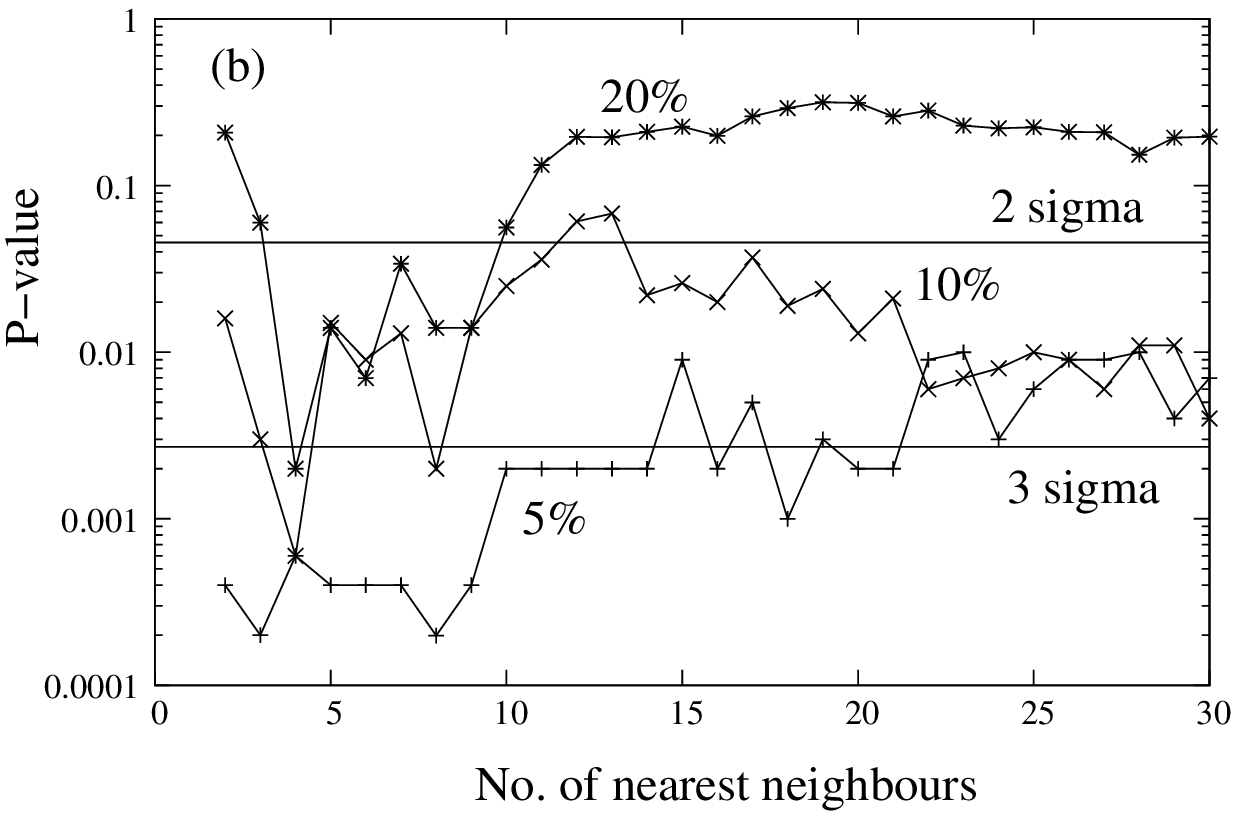}
    \caption{
 The P-value as a function of the number of 
nearest neighbours, $n_v$ using (a) Procedure I and (b) Procedure II. 
Here we consider data with error in polarization
flux less than 5\% (plusses), 10\% (crosses) and 20\% 
(crosses and plusses). 
}
\label{fig:error}
\end{figure}

In Fig. \ref{fig:SD_Polgt2} we show the $S_D$ values for 
data set with polarization flux greater than 2 mJy (set 2). 
The results for set 1 are shown for comparison. We find that
the values of $S_D$ do not change significantly between these two sets.
The resulting P-values for set 2 are shown in
Fig. \ref{fig:Pvalue_Polgt2}. We observe that 
the significance of alignment reduces. This reduced
significance can be explained by the fact that the number of samples
in this set is smaller in comparison to  
 set 1 by a factor of about 0.6. 
We should compare the results with these two cuts at fixed
 distance. 
Hence we should compare the results for $n_v = 10$ for set 1
 with $n_v=6$ for set 2. 
The mean value of the statistic 
of the random samples for a given $n_v$ is same for both the samples. 
We find that the mean is equal to 0.3668 and 0.3669 for $n_v=6$ for
set 1 and 2 respectively. 
The fluctuations for set 2 are larger than those for set 1 
by a factor
of $\sqrt{1/0.6}$ due to the smaller size of the sample. We find that
for $n_v=10$,
set 1, the standard deviation = 0.00495, whereas for $n_v=6$ and set 2,
 the standard deviation = 0.00698, consistent with our claim. This 
change in the standard deviation of the random samples explains the
difference in significance between the two samples. 

We next study the alignment  
further by imposing a cut directly on the
error in the polarization flux. In Fig. \ref{fig:error} we show the results
after imposing the cuts, error $\le 5$\%, $\le 10$\% and $\le 20$\% on data
using statistic $S_D$. 
The number of sources remaining after these cuts are 1058, 1728 and 2889  
respectively. 
The error in the polarization angle, in radians, is also
approximately equal to the fractional error in the polarization flux.
Hence these results can also be interpreted in terms of a cut
on the error in PAs. A 5\% error on the polarization flux leads to
a cut of approximately $3^o$ on the PA.  
 We clearly see that at small values of $n_v$, 
the data set with smallest error shows the highest 
significance. For example, data with error $\le 5$\% shows better
than 3 sigma significance for $n_v<10$. This set is clearly the most
reliable and provides further evidence that the effect we observe is
not caused by bias. In comparing with the results shown in Fig. \ref{fig:SD1}
we note that the sample size for error $\le 5$\% is much smaller,
roughly one-fourth of the sample used in Fig. \ref{fig:SD1}. 
For a homogeneous sample, the significance of alignment 
is expected to increase with
the sample size. Furthermore the physical distance for any value of $n_v$  
 in this sample would be roughly double the distance scale for same
$n_v$ for data set 1, which has 4400 sources. 
This shows that the
sample with error $\le 5$\% shows results with similar or better significance
in comparison to set 1.

\begin{figure}[!t]
    \includegraphics[width=5.0in,angle=0]{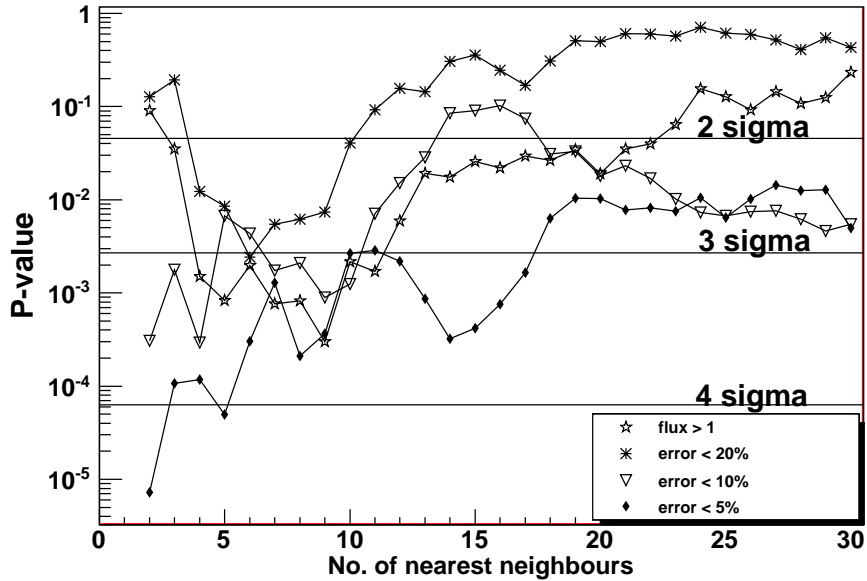}
    \caption{The
P-values 
for polarization flux $> 1$, error $\le 5\%$,
error $\le 10\%$ and error $\le 20\%$. Here
the random samples are generated using Procedure I. }
\label{fig:sigma_SDP}
\end{figure}

\begin{figure}[!t]
    \includegraphics[width=5.0in,angle=0]{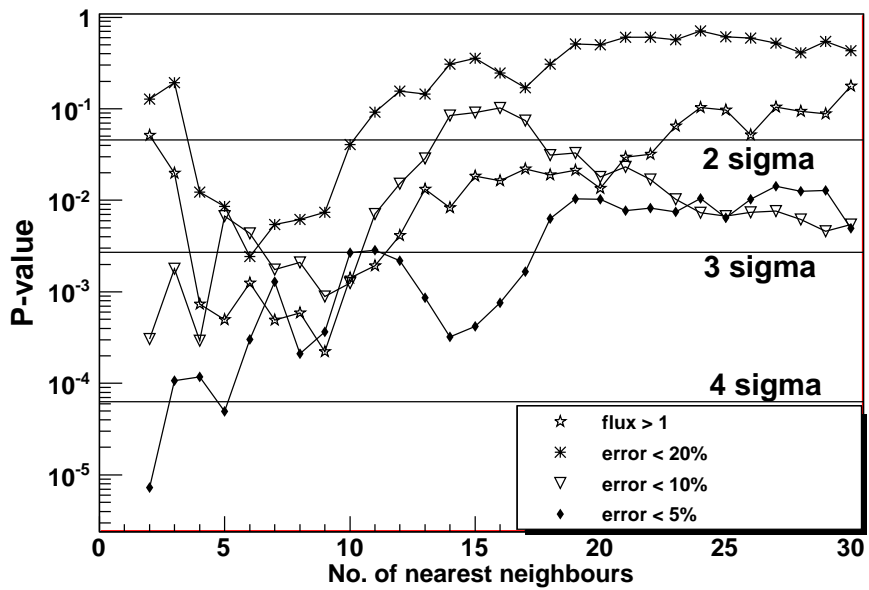}
    \caption{The
 P-values (lower graph)
for polarization flux $> 1$, error $\le 5\%$,
error $\le 10\%$ and error $\le 20\%$. Here
the random samples are generated using Procedure I. }
\label{fig:sigma_ZD}
\end{figure}

In Figs. \ref{fig:sigma_SDP} and \ref{fig:sigma_ZD} 
we show the results obtained using statistics 
$S_D'$ and $Z_D$ respectively for different cuts using Procedure I. 
Here the results are obtained by ignoring 
the contribution due to the parallel transport, which can be neglected for
small distances. We have explicitly verified that the results remain
practically unchanged irrespective of whether we include or exclude
it's contribution. We find that the results obtained with statistics
$S_D'$ and $Z_D$ are consistent with those obtained with $S_D$. In
general these statistics show higher significance in comparison to $S_D$.
We obtain similar results with these statistics if we use Procedure II.

\begin{figure}[!t]
    \includegraphics[width=5.5in,angle=0]{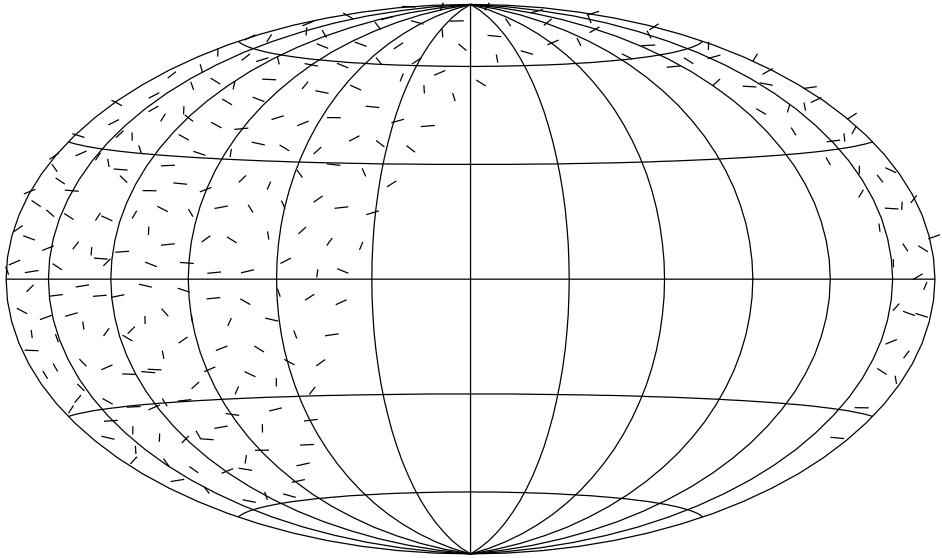}
    \caption{The linear polarizations averaged over a distance 
scale of 150 Mpc for
set $1$ in Galactic coordinates.
The small lines represent polarization angles
measured with respect to the local
longitude.  }
\label{fig:avg_pol}
\end{figure}

In Fig. \ref{fig:avg_pol} we show the polarizations for set 1 averaged
over 150 Mpc, the distance scale of alignment, in Galactic coordinates. 
Here we divide the celestial sphere by choosing a 
set of fixed latitudes in equatorial coordinates, which are further
subdivided to create equal area segments. 
This leads to a total of 273 segments. 
We show the average polarization angle in each region. 
The main purpose of this
plot is to visually detect any systematic pattern in polarization on
large angular scales. 
We observe that at low latitude, several polarizations 
appear to be locally aligned with the equator, i.e. have PA equal to
$\pi/2$ radians. This applies particularly to
sources lying at Galactic longitudes, $l \le \pi$ radians.
We are unable to detect any other striking pattern in Fig. \ref{fig:avg_pol}.

We next determine if the alignment of polarizations with Galactic latitude
is statistically significant.
Let $\chi_i$ represent the polarization angle
of source $i$ in Galactic coordinates. We define the statistic 
\begin{equation}
S_G = \sum_{i=1}^{n_s} \cos(2\chi_i - 2\chi_0)\big|_{\rm max}
\end{equation} 
The sum is maximized with respect to $\chi_0$. The resulting value
of $\chi_0$ is interpreted as the mean polarization angle. The corresponding
value of $S_G$ gives an estimate of how well the polarizations are 
aligned with the local longitude. A large value implies very good alignment.
We use this statistic to test for alignment of set 1 with the Galactic 
coordinates. 
We find that $S_G= 0.00721$. In this case also $n_s=4400$.
Comparing with 1000 random samples, the P-value in this case 
is 0.78. Hence we do not find significant alignment. We have 
repeated this analysis by including data only at very low
Galactic latitudes and also by keeping sources with $l\le \pi$ radians. 
The significance does rise in this case but 
remains much lower than 2 sigmas. 
We next apply this statistic by fixing the value of $\chi_0=\pi/2$. 
In this case again the entire data set does not show a significant signal. 
The significance for data at low Galactic latitudes in this case
is higher but still not equal to 2 sigmas. 
We do find a significant result at 2 sigmas if we also impose the cut
$l\le \pi$ radians. For $|b| \le 0.1$ radians and $l\le \pi$ radians, 
we find $P=0.03$ with the
number of sources, $n_s=77$. For $|b| \le 0.07$ radians and $l\le \pi$ radians, 
we obtain
$P=0.011$ with $n_s=36$. With futher reduction in $|b|$ the P-value
starts to rise. Hence we find a weak signal of alignment with 
Galactic equator at low latitudes. We do not understand the cause of
this alignment. However since the signal is relatively weak and obtained
only in a small subset of data, it might also be attributed to a 
statistical fluctuation. 
We have also determined how our results vary if we remove the region
$|b|< 10^o$. We do not find a significant change in results for any 
of the cuts discussed in this paper.

%\begin{figure}[!t]
%    \includegraphics[width=5.5in,angle=0]{Avg_PA_150_gal.eps}
%    \caption{The histogram of polarization in galactic coordinates for set 1,
%after averaging over a distance scale of 150 Mpc. }
%\label{fig:gal_pol_hist}
%\end{figure}

\subsection{Low Polarizations}

\begin{figure}[!t]
    \includegraphics[width=4.5in,angle=0]{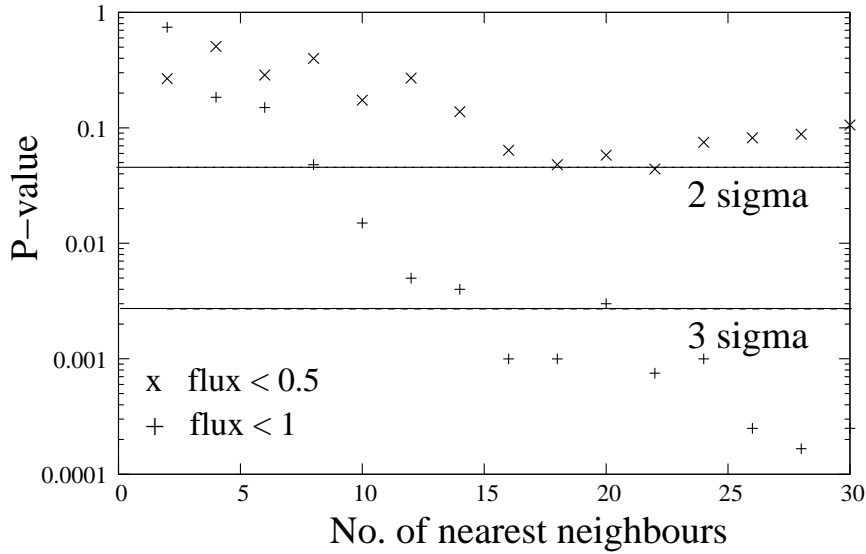}
    \caption{
 The P-value as a function of the number of 
nearest neighbours, $n_v$, for data  
set 3 (plusses)
and set 4 (crosses). 
}
\label{fig:pollt1.0}
\end{figure}

We have so far only considered results for data which is expected
to be most reliable. For comparison, we next show results for data
with low polarizations.
In Fig. \ref{fig:pollt1.0}, we show the significance of alignment for the
data set with polarization flux less than 1.0 mJy (set 3) and less than 0.5
mJy (set 4). 
For set 3 we find that P-value
is large for small $n_v$ and then rapidly decreases with increasing 
$n_v$. This trend is exactly opposite to that seen in 
Fig. \ref{fig:SD1} for 
data set 1. 
The fact that data with low polarization flux
shows no signal of alignment at small $n_v$ provides further evidence that
the alignment seen in set 1 cannot be attributed
to bias. This is because a bias will dominantly affect data with low
polarization flux and any signal which arises due to bias should be more 
prominent in this set. This point is further reinforced by data 
corresponding to polarization flux less than 0.5 mJy. In this case we do
not find significant alignment over the entire range of $n_v$ shown in
 Fig. \ref{fig:pollt1.0}.
      
\begin{figure}[!t]
\hskip 0.1in
    \includegraphics[width=4.5in,angle=0]{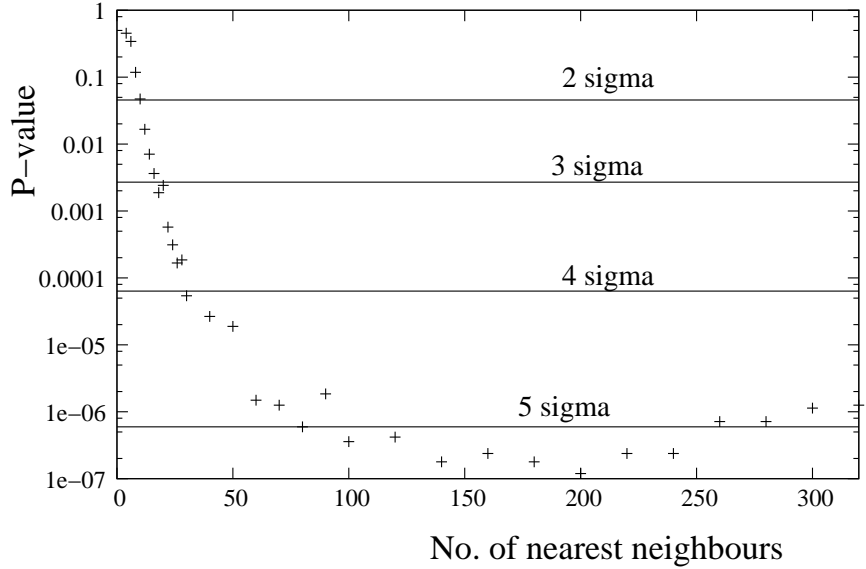}
    \caption{
The P-values  
as a function of the  number of
nearest neighbours, $n_v$, for set 3.
}
\label{fig:pollt1.0_1}
\end{figure}

\begin{figure}[!t]
    \includegraphics[width=4.5in,angle=0]{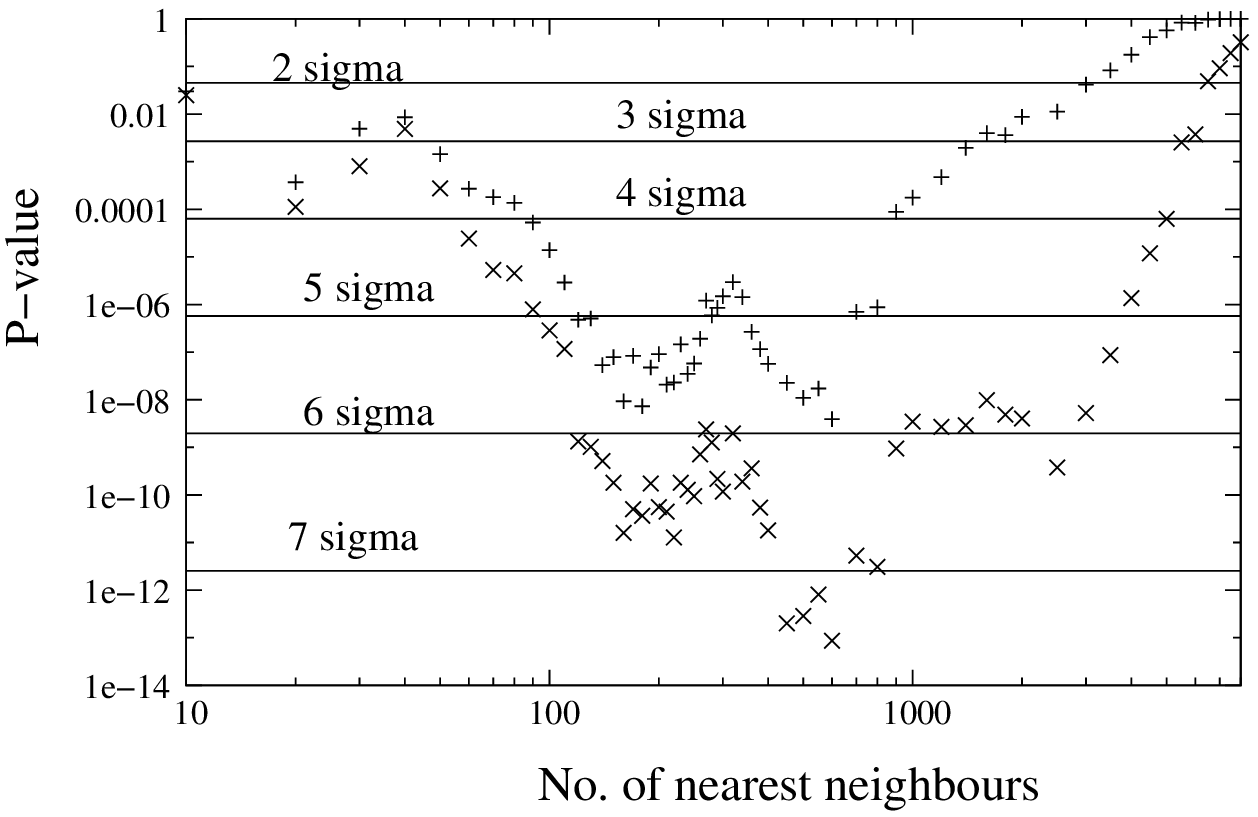}
    \caption{The P-value for set 3 using Procedure I (plusses) 
and II (crosses)
as a function of $n_v$.
Here we have used statistic $S'_D$.  
}
\label{fig:Pollt1_1}
\end{figure}

\begin{figure}[!t]
    \includegraphics[width=4.5in,angle=0]{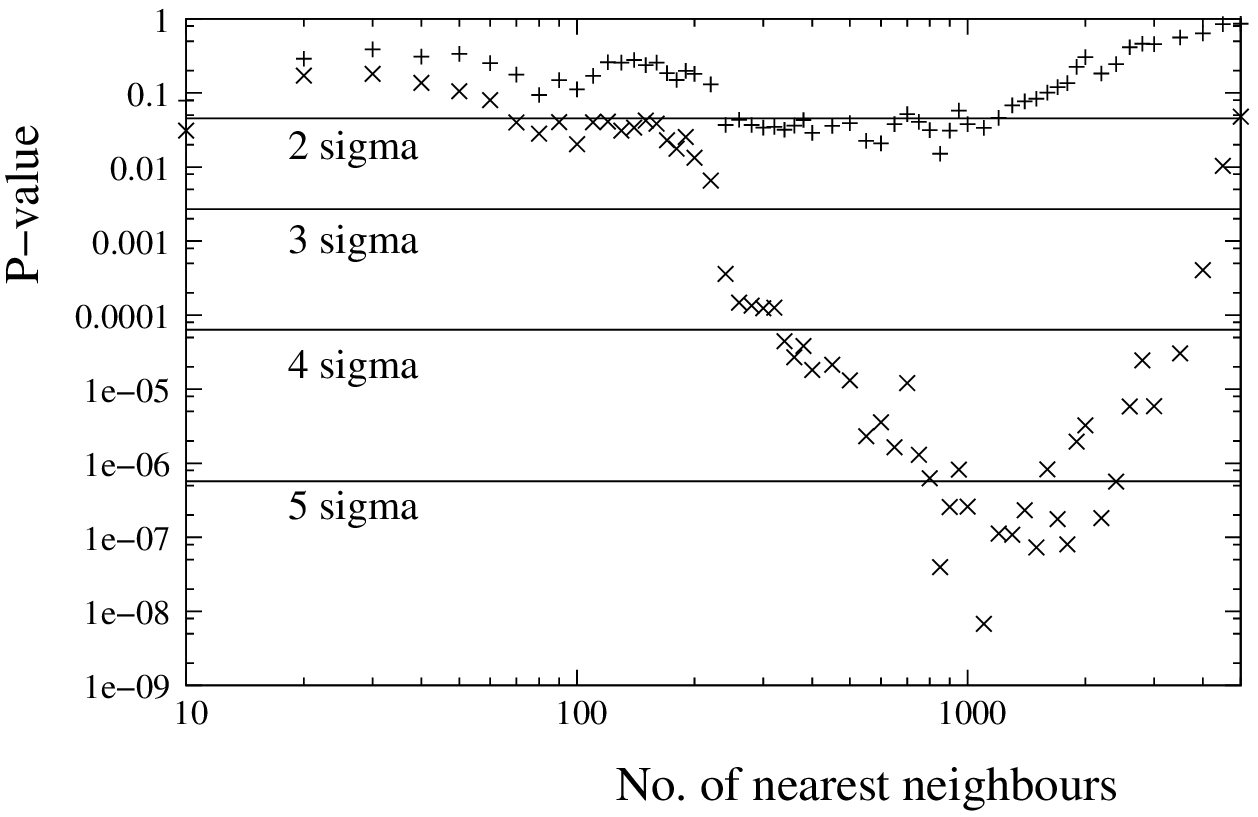}
    \caption{The P-value for set 4 using Procedure I (plusses) and II (crosses)
as a function of $n_v$.
Here we have used statistic $S'_D$.  
}
\label{fig:Pollt0.5_1}
\end{figure}

In Fig. \ref{fig:pollt1.0} we observe a 
high significance of alignment for $n_v>15$ for data set 3. 
This data set, however, has higher error in comparison to 
set 1. For this reason we have confined
ourselves mostly to set 1. One may speculate that the high significance
we see in this set at larger distances arises due to bias. However 
this interpretation is not consistent since set
 4 shows no effect. The bias would have affected this sample the most. 
We add that we have studied set 4 at larger values of $n_v $ and found no 
signal of alignment with Procedure I. 
In contrast set 3 continues to yield a significant
signal for larger $n_v$, as shown in 
Fig. \ref{fig:pollt1.0_1}.  
Here the P-values are found to be very small for some $n_v$.
Hence we compute them 
by using the sigma values.
Here we
see that for set 3, the significance continues to increase 
upto $n_v=200$. This corresponds
to a distance scale of about 500 Mpc. 

In Fig. \ref{fig:Pollt1_1} we show the P-value over a much larger range
for set 3. The corresponding result for set 4 is shown in 
Fig. \ref{fig:Pollt0.5_1}. Here we use the statistic $S'_D$ instead of $S_D$. 
This is because $S_D$ takes very long computation time for large values
of $n_v$. We have verified that the two statistics give similar results in all
cases where the results for both have been obtained. This can also be seen
by comparing the results of Fig. \ref{fig:sigma_SDP} with those of  
Figs. \ref{fig:error}a and \ref{fig:SD1}. The results of
 Fig. \ref{fig:Pollt1_1} are also consistent with those obtained
with statistic $S_D$, Fig. \ref{fig:pollt1.0_1}, where we find the 
most significant result at $n_v\approx 200$. However at larger $n_v$ we find an additional dip in the P-value
at $n_v\approx 600$. This corresponds to a distance scale of about 850 Mpc. 
Beyond $n_v=600$, the significance reduces monotonically.  
A very interesting feature of Fig. \ref{fig:Pollt1_1} is the difference
in results seen with Procedure I and II. We find that Procedure II
in general yields higher significance. The difference continues to
increase until $n_v\approx2500$ and then starts decreasing.
We observe a very high significance at $n_v\approx2500$ with Procedure II. 
At this $n_v$, Procedure I yields a null result. This indicates that
the polarizations have a tendency to point in the same direction over
very large distances. Procedure I does not yield significant result
at $n_v=2500$ since in this case the random samples will also have this
correlation. 
 This trend
is very similar to that seen in Fig. \ref{fig:Pvalue_noshuffle} at large
$n_v$. 
It is possible that this very large scale correlation
arises due to instrumental bias since it affects almost the entire data.
This is further indicated by the results for set 4, seen in Fig. 
\ref{fig:Pollt0.5_1}. Here we find that Procedure I does not yield a
significant result for any value of $n_v$. However Procedure II yields a
very strong signal at large $n_v$. The strongest signal is seen at $n_v\approx
1100$ and is similar to the trend seen in set 3 at $n_v\approx 2500$.  
We discuss this issue in further detail in Section 5.

We find a similar trend with cuts on degree of 
polarization ($dop$). For $n_v > 40$, the 
data with $dop>1$ \% shows alignment whereas
the remaining set does not.  
The complete data set of 12743 sources yields 
a $5.4$ sigma signal of alignment at $n_v=250$, 
which also corresponds to a distance scale of order 500 Mpc.
The signal decays below 3 sigma
only for $n_v>750$.

The results seen for set 3 and 4 using Procedure I 
can be explained by realizing that 
a physical effect of alignment is likely to be masked in a sample which
has large error. For larger values of $n_v$ the fluctuations get 
reduced since each source is compared with a larger number of nearest 
neighbours. Hence the significance is expected to get enhanced. 
This is precisely the effect seen in data.  
We point out that sets 1 and 2 also show a similar trend with low
significance for very small $n_v$.
In the case of set 4, the error is very large. This is most likely
the reason for the lack of signal seen in this set.

Due to the presence of relatively large error in low polarization data set,
we suggest that these results be tested further with 
more refined observations. If these results are confirmed with more 
refined data, then it would suggest the existence of correlations
on larger distance scale for data with low polarization flux. 

\section{Bias}
\label{sc:bias}
The JVAS/CLASS data may contain bias due to error in the removal
of residual instrumental polarization \cite{Jackson:2007,Joshi:2007}. 
In different observation sessions, the  observations 
were made on different regions of the sky. Faulty removal of
instrumental bias could lead to large scale correlations in these
regions. We expect that a bias
might affect the low polarization sample the most. 
It is possible that the bias is so large that all the data set is
significantly affected. In that case we cannot draw any conclusions
from the data. Here we assume that the bias is relatively small.
 The fact that set 4
does not lead to a significant signal of alignment gives us considerable
confidence that bias is not responsible for the correlations we observe using
Procedure I.
It is possible that the non-uniformity in distribution, 
i.e. an excess of
sources with PAs lying between $0^o$ and $90^o$, 
seen 
in Fig. \ref{fig:PAs}, might be caused by instrumental bias. 
This non-uniformity 
indicates the presence of some systematic effect which affects the
data globally.
Furthermore it is roughly correlated with the equatorial
coordinate system with the shift occurring close to $90^o$, which also
suggests that it might be caused by bias. 
Alternatively it might arise due to some local effect, caused, for example, 
by the ionosphere.
As we discuss below, any such bias, which might affect the entire data, 
cannot generate the correlations we report.
Furthermore there is evidence in the data which suggests that 
the non-uniformity
in the distribution, irrespective of its origin, 
affects our results primarily at very large $n_v$,
where the significance of the signal observed with 
Procedure I is much reduced.

In our analysis we have mostly used Procedure I which generates random
samples by shuffling real PAs among different sources. In this case each
random sample has the same PA distribution as the real data.
This procedure naturally eliminates some contributions which might 
arise due to bias. In particular, a bias which might affect the entire 
sample in the same manner would not yield a significant signal of
alignment with Procedure I.  
Let us consider the extreme case where 
the polarization angles in the entire data set are biased in the same manner,
i.e. they point in the same direction, upto some contribution due to 
random error.
In this case the distribution of PAs will be
maximally skewed. 
This bias
cannot yield the alignment signal we observe using Procedure I since
 the bias will affect the random samples as much as it
affects real data. The random samples will also have the same 
PA values at all sites as the real sample. The difference between the 
two will arise primarily 
due to random error. 
The statistic of random samples will be exactly equal to that of real sample
upto fluctuations due to random error in PAs.
Hence we do not expect to get a significant signal in this case.  

In the present sample it is more likely that different regions of the 
sky might be biased differently. Let us assume that this bias acts over
very large distance scales. In that case it is reasonable to assume that
the non-uniformity in the distribution, seen in Fig. \ref{fig:PAs}, is 
caused by this bias. It is, therefore, interesting to determine the influence
of this non-uniformity on our results.
As we have already explained in section 3, 
this would affect the results of Procedure II but not
Procedure I, which preserves the distribution of the real data
while generating random samples.
The difference in results
between Procedures I and II, therefore, 
provides us with an estimate of the effect
caused by the non-uniformity in distribution. 
This difference can be seen
in Figs. \ref{fig:Pvalue_noshuffle}, \ref{fig:Pollt1_1} and
\ref{fig:Pollt0.5_1} for sets 1, 3 and 4 respectively. 
It is relatively small
for small $n_v$ with Procedure I giving a smaller significance. 
As discussed in section 4, the
main difference comes at large $n_v(\sim 2000)$ for set 1. Here Procedure I
yields an   
anomalously low significance (high P-value) and, in contrast, Procedure II 
yields a mild (2 sigma) evidence of correlation. The strongest difference
between the two procedures is seen at $n_v\approx 2000$, almost half of 
the entire sample. This indicates the presence of a systematic effect
which affects nearly half the sample in the same manner. This trend is
seen even more prominently in sets 3 and 4. Here Procedure II yields
very significant result ($\sim 6 \sigma$) at $n_v=2500$ and 1100 respectively, whereas 
Procedure I yields a null result at these $n_v$ values. 
For set 1,
Procedure II shows a dip in P-value at $n_v=2000$ since the random 
samples in this case are free from the systematic effect present in 
data. In contrast, the random samples corresponding to Procedure I
will also have such
a systematic component. Hence the possibility that these 
might yield a larger statistic than the real data is not unlikely.  
This argument also explains the results seen in set 3 and 4.
We conclude that 
the non-uniformity in distribution only affects our results at
very large $n_v$ and does not generate the correlations that we report with
Procedure I.

Finally, we address the issue of bias raised in Battye {\it et al}\cite{Battye:2008}. The
authors found a significant bias present in the polarization angles in the
NVSS survey. It was found that the polarization angles have a tendency to
be close to multiples of $45^o$. 
As can be seen from Fig. \ref{fig:PAs}, this bias is absent for all
the data sets we study in this paper. We do find a peak close to $45^o$
but no peaks at $90^o$, $135^o$ and $180^o$. Hence, although the distribution
shows deviation from uniformity, the trend is different from that discussed
in Battye {\it et al}\cite{Battye:2008}. 

Here we have argued that it is unlikely that the correlations we observe
with Procedure I
arise due to bias in the data. Specifically we have examined the
effect of a bias which might affect the data globally. We have shown
that it does not affect our results corresponding to Procedure I. 
Further evidence for absence of bias comes from the fact that data with
low polarizations (set 4) shows no effect. 
However we cannot rule out the 
possibility of bias entirely and the 
issue may be best resolved by more
refined data.

\section{Physical Explanation}
In this section we provide a possible physical explanation for the 
results obtained in this paper. We first notice that the alignment observed
in set 1 (Pol. flux $>$ 1 mJy) is over a distance scale of order 150 Mpc.  
At such distances the Universe does not show homogeneity and isotropy.
For example, we
observe superclusters of galaxies with length scales of order 100 Mpc.
Hence we cannot rule out correlations among galaxies at
such distances within the framework of the Big Bang cosmology. 
The distance scale observed also agrees well with the observed peak 
in the  
galaxy correlations \cite{Eisenstein:2005} using Sloan Digital Sky 
Survey, as predicted by the Big Bang cosmological model. The peak
indicates an enhancement in the galaxy-galaxy correlations at a distance
scale of 150 Mpc.  
We still 
require a physical mechanism which would cause such an alignment. 
One possibility is the large scale correlation in the intergalactic 
magnetic field \cite{Agarwal:2011,Agarwal:2012}. 
If the intergalactic 
magnetic field has a primordial
origin \cite{Subramanian:2003sh}, then it could show correlations over 
large distance scales. 
Such correlations could generate
the observed alignment by either affecting the electromagnetic radiation
during propagation through intergalactic medium or by intrinsically
aligning the sources at high redshift. 
We further speculate that correlations in the intergalactic magnetic field
might also show enhancement at the scale of 150 Mpc, as observed 
in the galaxy correlations \cite{Eisenstein:2005}. This might 
explain the strong alignment seen at this distance scale, corresponding
to $n_v=10$, as shown
in Fig. \ref{fig:SD1}.

\begin{figure}[!t]
    \includegraphics[width=4.5in,angle=0]{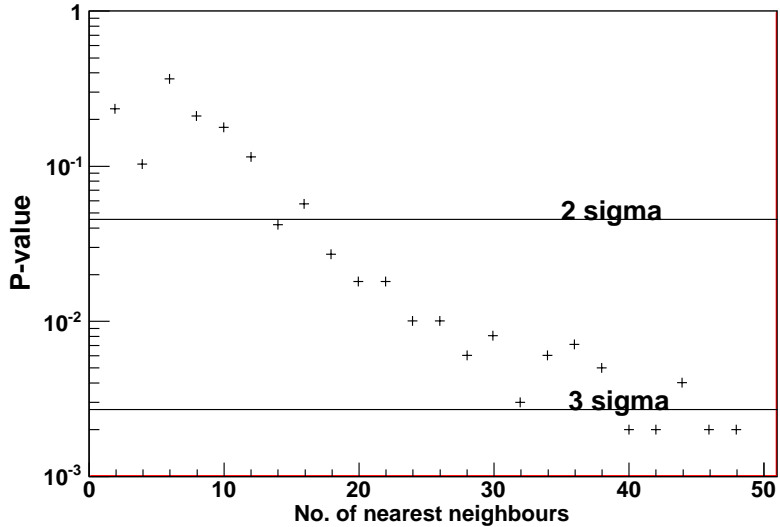}
    \caption{The P-value for a simulated low polarization sample which has
large error. The sample has correlations over 
large distances, as described
in text. As expected, we see that the significance of alignment increases
with the number of nearest neighbours.
}
\label{fig:pol_sim}
\end{figure}

The correlation seen in the low polarization sample (set 3) is explained
by postulating correlations in the magnetic field at larger distances
of the order of 500 Mpc. 
The low polarization sample does not show alignment at short
distances since the error in this sample is much larger. Hence
the effect becomes apparent only when we compare a particular source
with a sufficiently large number of nearest neighbours. In this case
the fluctuations in the estimate of $d_k$ at any point get reduced. 
We demonstrate this phenomenon with an explicit simulation. We 
divide the whole sphere in 12 equal area parts using 
HEALPix\footnote{http://healpix.jpl.nasa.gov/}. We generate a total of 3072 
sources distributed uniformly over the sphere. For each segment the
source PAs are generated from a Gaussian distribution with fixed mean and
 large standard deviation, equal to $60^o$. The standard deviation
is taken to be large in order to simulate the large error in real PA 
values. The mean value in different segments is chosen randomly.  
We next determine the significance of alignment in this data set, considering
only the sources in the 
 northern hemisphere. The resulting P-values are shown in Fig. \ref{fig:pol_sim}.  
We observe that at very low $n_v$, the correlation is not significant.
However we see a strong signal  for large $n_v$, in agreement with
what is observed in set 3. This also explains the trend seen in set 1  
and 2 and in optical data \cite{Hutsemekers:1998} 
where a significant signal is not seen at very low $n_v$. 
The significance
gets enhanced for larger values of $n_v$. 

The high polarization sample (set 1) does not show alignment signal at
distance scale of order 500 Mpc since these sources may have evolved
due to galaxy clustering. 
Hence these show correlations only 
upto the clustering distance scale. 
The magnetic field of these sources would
be influenced by the cluster magnetic field and hence is likely to loose
coherence with sources at larger distances.
We speculate that sources which have low 
 polarizations may not be significantly
affected by clustering of galaxies. This may be another reason why
these sources do not show an effect at short distances.

The above explanation for low polarizations 
requires that magnetic field shows correlations
on large distances of the order of 500 Mpc.
Such correlations may be present in the  
 primordial model of intergalactic magnetic field \cite{Subramanian:2003sh}. 
Hence one may be able to explain these also within the framework
of Big Bang cosmology. However  
 a more detailed analysis is required to verify this possibility, which
we postpone to further research.

It is natural to expect that both the optical and radio alignment 
might have a common origin, which may be the intergalactic magnetic field.
However, in detail, these two may show different behaviour since
there are many physical effects, such as extinction, mixing with 
hypothetical pseudoscalar particles,
which affect optical but not radio frequencies. A possible scenario is that 
the radio alignment arises due to intrinsic alignment of galaxies, 
generated by the large distance correlations in the 
intergalactic magnetic field. It is generally expected 
that primordial magnetic field provides the seed for
the magnetic field of these sources. The linear polarization of radio waves
from these sources 
would be directly correlated with the direction of their magnetic field
 and may lead to
the observed alignment in radio polarizations. 
The optical alignment, however, might arise dominantly due to 
propagation. For example, it might acquire 
correlations over larger distance due to
 mixing with pseudoscalar particles in background
magnetic field \cite{Agarwal:2011,Agarwal:2012}.  
This phenomenon  
 does not significantly affect 
radio polarizations \cite{Payez:2012pm}.  
Hence despite the observed difference in the scale of alignment, it is
possible that intergalactic magnetic field might explain both observations.

Our proposal that polarization alignment arises due to intrinsic
alignment of sources may be tested by study of the corresponding jet angles.
Information about jet angles is available for a small subsample 
of our data set \cite{Joshi:2007}. The histogram of the difference
between the jet PA 
and the polarization PA shows a peak at $90^o$ 
\cite{Joshi:2007}. Due to this correlation between the two PAs,  
the jet angles may also show alignment seen in polarization PAs.
For set 1, we are able to unambiguously find only 520 sources for which 
information about the jet angles is also available. We have tested
this sample for alignment at small $n_v$. It 
does not show a significant signal either for polarization PA or jet angles.  
This null result might arise since this sample
is too small and hence the signal of alignment would be weak. Furthermore
even for $n_v=2$, the mean distance between galaxies is larger than 150 Mpc.
Beyond this distance the correlations in set 1 decay rapidly. 
Hence the results of this study of jet angles is inconclusive.
We suggest that this proposed explanation should be tested further
by a more detailed survey. 

If our proposed explanation in terms of intergalactic magnetic field
is applicable, then it might imply that sources which are dominated
by local effects might not show the alignment effect. These may 
be sources which show relatively strong magnetic field, which
are generated during their evolution. In order to study this possibility
we eliminated the sources with
degree of polarization ($dop$) greater than $10$ \% in set 1. We find that
this leads to an enhancement of the signal at $n_v=2$. The P-value now reduces
to $0.04$ instead of $0.11$ seen in the entire sample. For larger values
of $n_v$ we do not see a clear trend. On average we find a larger P-value
which may arise due to reduced number of sources. It is likely 
that local effects would be most prominent for smaller values of $n_v$.
Hence our results give a suggestive but rather weak evidence that local effects
may destroy coherence in sources. These issues require 
further study, with more refined data, 
which we postpone to future research.

%\section{Discussion}
\section{Conclusions}
\label{sc:conclusion}
%%%%%%%%%%%%%%%%%%%%
The possibility that sources at high redshift might show 
correlations at very large distance scales was first indicated
by \cite{Hutsemekers:1998}. Here it
was found that optical polarizations from quasars show alignment
over very large distances. A similar test at 
 radio frequencies was first investigated
in \cite{Joshi:2007}, who found a null result. 
 Here we test for alignment in radio
polarizations using a coordinate invariant statistics. We 
 confirm the results of \cite{Joshi:2007}, obtained in data which
includes only sources with 
 polarization flux greater than 1 mJy. However they did not
explore the possibility of alignment at small distances. 
We find a significant signal for
small number of nearest neighbours. The
distance scale of alignment here is found to be of order 150 Mpc.
At such distances,
galaxies are expected to show clustering
and
 the Universe is not expected to be homogeneous
and isotropic.
Hence the results we find are not inconsistent with 
Big Bang cosmology.

At larger distances the data set with polarization flux greater than 1 mJy
does not yield a significant signal of alignment. However, surprisingly,
we find that this set shows unusually large scatter.
The significance of alignment tends to fall
below $-2\sigma$ at large distance.  
This anomalous behaviour is seen only when comparing with respect to 
randomly shuffled data. It is not observed if 
the random data set is generated from 
a uniform distribution. We argue that this anomalous behaviour most likely
arises since the distribution of observed polarizations is mildly non-uniform.
It is possible that this non-uniformity may be caused by bias. However 
we have shown that it
has not effect on our results, as long as we use Procedure I.  
In any case, due to the presence of such anomalies, 
it is necessary to test our 
results with more refined data.

We also find that data with polarization flux less than 1 mJy does not
show a significant signal for short distances or, equivalently, 
small number of nearest
neighbours. If the signal was caused by bias, then it would have been 
dominant in this set. This provides us considerable confidence that the
signal of alignment we observe is likely of physical origin. 
We find a similar trend if we impose cuts on data based on the error
in polarization flux. The significance of alignment in data with low
error is found to be large at short distances and the effect disappears
at large distances. We also find that the significance at small distances
becomes better as the error becomes smaller. 

The low polarization data, with polarization flux less than 1 mJy, 
as well as the complete data set, does, however, 
show a very significant signal 
alignment at larger distances. The strongest signal 
is seen at distances of order 500-850 Mpc with significance better than
$5\sigma$. 
Although this set has larger error, it is unlikely that this signal is
caused by bias. This is because the sample with polarization flux less
than 0.5 mJy does not show significant correlations for any value of $n_v$
with Procedure I.
A bias would have affected this set the most. 
This signal might indicate the presence of correlations in intergalactic
magnetic field at  
distance scales larger than the scale of galaxy clusters. 
Due to the presence of large error in the low polarization sample,
 more refined data is required in order to properly 
study this effect.

The signal of alignment we find is similar but not  
identical to that found by 
\cite{Hutsemekers:1998} for optical polarizations. 
There the signal of alignment was found over cosmologically large distances
whereas here the 
alignment is seen only over somewhat smaller 
distances
of order 150 Mpc for the sample with polarization flux greater than 1 mJy.  
The sample with polarization flux less than 1 mJy, however, 
shows alignment for distances of order 500 Mpc.

We have also suggested a physical model which may explain our results.
 The large scale alignment of radio
polarizations is explained by assuming that the background intergalactic
 magnetic field, which may be of primordial origin, has large scale
correlations in real space \cite{Subramanian:2003sh,Seshadri:2005aa,Seshadri:2009sy}. This magnetic field provides the seed for
the magnetic field of the radio sources. Hence some large distance
correlations may be present in magnetic field of these sources also.
The radio polarizations would be directly aligned with the source magnetic
field and hence show large distance alignment.

%%%%%%%%%%%%%%%%%%%%%%%%%%%%%%%%%%%%%%%%%%%%%%%%%%%%%%%%%%%%%%%%%%%%%%%%%%%%%%%%%%%%%%%%%%%

%%%%%%%%%%%%%%%%%%%%%%%%%%%
\section*{Acknowledgements}
We have used CERN ROOT 5.27 for generating our plots. Prabhakar Tiwari
sincerely acknowledge CSIR, New Delhi for award of fellowship during the work. 
%%%%%%%%%%%%%%%%%%%%%%%%%%%
\bibliographystyle{ws-ijmpd}
\bibliography{radio}
\end{document}